\documentclass[10pt,english]{article}
\usepackage[T1]{fontenc}
\usepackage[latin9]{inputenc}
\usepackage{color}
\usepackage{setspace}
\usepackage{babel}

\begin{document}
\begin{singlespace}

\title{\noindent \textcolor{black}{Minimum accelerations from quantised
inertia.}}
\end{singlespace}

\begin{singlespace}

\author{\noindent \textcolor{black}{M.E.~McCulloch }%
\thanks{\textcolor{black}{Marine Science \& Eng., University of Plymouth,
UK. mike.mcculloch@plymouth.ac.uk}%
}\textcolor{black}{ \  }%
\thanks{\textcolor{black}{School of Physics, University of Exeter, UK, M.E.McCulloch@exeter.ac.uk}%
}\textcolor{black}{ }}
\end{singlespace}

\maketitle
\begin{singlespace}

\section*{\noindent \textcolor{black}{Abstract}}
\end{singlespace}

\begin{singlespace}
\noindent \textcolor{black}{It has recently been observed that there
are no disc galaxies with masses less than $10^{9}M_{\odot}$ and
this cutoff has not been explained. It is shown here that this minimum
mass can be predicted using a model that assumes that 1) inertia is
due to Unruh radiation, and 2) this radiation is subject to a Hubble-scale
Casimir effect. The model predicts that as the acceleration of an
object decreases, its inertial mass eventually decreases even faster
stabilising the acceleration at a minimum value, which is close to
the observed cosmic acceleration. When applied to rotating disc galaxies
the same model predicts that they have a minimum rotational acceleration,
ie: a minimum apparent mass of $1.1\times10^{9}M_{\odot}$, close
to the observed minimum mass. The Hubble mass can also be predicted.
It is suggested that assumption 1 above could be tested using a cyclotron
to accelerate particles until the Unruh radiation they see is short
enough to be supplemented by manmade radiation. The increase in inertia
may be detectable.}
\end{singlespace}

\begin{singlespace}

\section{\noindent \textcolor{black}{Introduction}}
\end{singlespace}

\begin{singlespace}
\noindent \textcolor{black}{It was first noticed by {[}1{]} that galaxies
were too energetic to be held together by their visible matter and
he proposed the existence of an invisible (dark) baryonic matter that
provides the extra required gravitational pull. Dark matter is now
assumed to be non-baryonic because of the tight constraints on existing
baryonic matter from cosmological nucleosynthesis, and it is still
the most popular explanation {[}2{]} for the galaxy rotation problem
{[}3{]}, {[}4{]}. However, after decades of searching, dark matter
has not been directly detected, though attempts to do so are ongoing,
for example: DAMA/Libra {[}5{]}, CDMS-II {[}6{]}, and XENON10 {[}7{]}
and intercomparisons {[}8{]}.}

\noindent \textcolor{black}{An alternative explanation for the galaxy
rotation problem was proposed by {[}9{]} who suggested that either
1) the force of gravity may increase or 2) the inertial mass ($m_{i}$)
may decrease in a new way for the very low accelerations at a galaxy's
edge. His empirical scheme (Modified Newtonian Dynamics, MoND) fits
galaxy rotation curves, and has the advantage of being less tunable
than dark matter. However, it does require a tunable parameter, the
acceleration $a_{0}$, and a tunable interpolation function, and it
does not correctly model galaxy clusters. MoND is also a controversial
theory since, in its gravitational variant it disagrees with Solar
System dynamics for some values of its interpolation function {[}10{]},
and in its inertial variant it violates the equivalence principle
(although this is a lesser problem since this principle has not been
tested for the low accelerations seen at the edges of galaxies, which
are unobtainable on Earth {[}11{]}.}

\noindent \textcolor{black}{The Pioneer anomaly is similar to the
galaxy rotation problem. It is an apparent anomalous acceleration
towards a gravity source, in this case the Sun, of $8.7\times10^{-10}m/s^{2}$
and was first observed by {[}12{]} and {[}13{]}. It has not been detected
in the acceleration of the planets and some of their satellites {[}14{]},
{[}15{]}, {[}16{]}, {[}17{]}, {[}18{]}, {[}19{]}, {[}20{]}, excluding,
perhaps, Neptune {[}21{]}, Pluto {[}22{]} and trans-Neptunian objects
({[}23{]} and {[}24{]}) whose trajectory data may not yet be accurate
enough. This implies that the explanation is not a simple modification
of gravity, since this would also effect the planets.}

\noindent \textcolor{black}{The possibility that the Pioneer anomaly
is a consequence of an anisotropic dissipation of onboard sources
of thermal radiation has been investigated by {[}25{]}, {[}26{]},
{[}27{]} and {[}28{]}. They have found that this process could explain
up to one third of the anomaly, but it is difficult to see how this
could explain the constant anomalous acceleration, and the possible
onset of the anomaly at 10 AU.}

\noindent \textcolor{black}{Other suggestions have included an empirical
modification of gravity {[}29{]}, and the addition of an extra dimension
within General Relativity {[}30{]}. Many of these suggestions are
summarised in {[}31{]} and a space mission to test for the anomaly
was recently suggested by {[}32{]}.}

\noindent \textcolor{black}{In summary the anomalous acceleration
seems to be non-gravitational, dependent on trajectory (the bound
planets are unaffected at the same orbital radii), and so it makes
sense to investigate inertia {[}33{]}, also because it is not well
understood.}

\noindent \textcolor{black}{Following the prediction of Hawking radiation
from black holes {[}34{]}, {[}35{]} suggested that an accelerating
body may also see thermal radiation and {[}36{]} derived an inertia-like
force from this Unruh radiation. The wavelength of Unruh radiation
increases as the acceleration reduces and {[}37{]} pointed out that
at the very low accelerations at which galaxies start to deviate from
expected behaviour, the Unruh wavelength reaches the Hubble scale.
He suggested that then the Unruh waves may not be observable, and
that somehow this may abruptly reduce the Unruh-inertia at these low
accelerations. This idea is suggestive, but does not quantitatively
fit any galaxy rotation curves, and cannot explain the Pioneer anomaly,
since the Unruh waves these spacecraft may see are far shorter than
the Hubble scale.}

\noindent \textcolor{black}{Building on Milgrom's abrupt break, a
new model for inertia was proposed by {[}38{]} and this model could
be called a Modification of inertia resulting from a Hubble-scale
Casimir effect (MiHsC) or Quantised Inertia. As above, MiHsC assumes
that the inertial mass of an object is caused by a drag from Unruh
radiation. The new assumption is that this Unruh radiation is subject
to a Hubble-scale Casimir effect. This means that only Unruh wavelengths
that fit exactly into twice the Hubble scale (harmonics with nodes
at the boundaries) are allowed, so that a greater proportion of longer
Unruh waves are disallowed, reducing inertia in a new, more gradual,
way for low accelerations. This model predicts that the inertial mass
($m_{I}$) varies as}

\noindent \textcolor{black}{\begin{equation}
m_{I}=m_{g}\left(1-\frac{\beta\pi^{2}c^{2}}{\left|a\right|\Theta}\right)\end{equation}
}

\noindent \textcolor{black}{where $m_{g}$ is the gravitational mass,
$\beta=0.2$ (empirically derived by Wien for Wien's law), c is the
speed of light, $\Theta$ is the Hubble diameter ($2.7\times10^{26}m$,
{[}39{]}) and the acceleration (a) for the Pioneer craft was the magnitude
of the acceleration of the Pioneer 10 and 11 spacecraft relative to
their main attractor the Sun (see {[}38{]} for the derivation of Eq.
1). MiHsC predicted a reduction of inertial mass for the Pioneer craft
of about 0.01\% which made them more sensitive to the Sun's gravity
resulting in an extra predicted Sunward acceleration of $6.9\times10^{-10}m/s^{2}$,
in agreement with the Pioneer anomaly. The advantages of MiHsC are
that it is based on a physical model, is simple, and it needs no adjustable
parameters. Its main disadvantage is its apparent agreement only with
unbound trajectories.}

\noindent \textcolor{black}{{[}40{]} showed that MiHsC agrees quite
well with the flyby anomalies: unexplained velocity changes seen in
Earth flyby craft observed by {[}41{]} (also discussed in: {[}42{]}
and {[}43{]}), if the conservation of momentum is considered, and
also if the acceleration in Eq. 1 is not the acceleration with respect
to a single background (as in Milgrom's MoND or {[}44{]}), but instead
the acceleration relative to the surrounding matter (following Mach's
principle). In {[}45{]} this was taken to its logical conclusion,
and the acceleration relative to the fixed stars was included in Eq.
1.}

\noindent \textcolor{black}{Returning to the galaxy rotation problem:
this is also an unexpected acceleration towards a source of gravity.
The problem is that galaxies are bound systems and MiHsC only seems
to apply to unbound objects. This problem is avoided here by looking
at galaxies at the edge of boundedness, ie: very low mass ones in
the hope that their characteristics may be determined by MiHsC. }

\noindent \textcolor{black}{Recently, {[}46{]} studied the baryonic
mass of rotationally supported (disc) galaxies and showed that there
are none with a baryonic mass less than $10^{9}M_{\odot}$ (within
the central 500 parsecs) and these results are unexplained so far.
In this paper it is shown that MiHsC predicts minimum linear and rotational
accelerations. The former are shown to be close to the acceleration
attributed to dark energy, and the latter imply a minimum disc galaxy
mass close to that observed.}
\end{singlespace}

\begin{singlespace}

\section{\noindent \textcolor{black}{Method \& Results}}
\end{singlespace}

\begin{singlespace}
\noindent \textcolor{black}{Starting with Newton's second law and
gravity laws for a star with gravitational mass $m_{g}$ and inertial
mass $m_{I}$ orbiting a galaxy with gravitational mass $M$ we get
\begin{equation}
F=m_{I}a=\frac{GMm_{g}}{r^{2}}\end{equation}
Following {[}38{]},{[}40{]} and {[}45{]} the inertial mass is replaced
with Eq. 1, in which |a| is the average mutual acceleration of every
other mass in the universe. Simplifying the constants ($\beta\pi^{2}\sim2$,
introducing a 1.5\% error) and cancelling $m_{g}$ we get}

\noindent \textcolor{black}{\begin{equation}
\left(1-\frac{2c^{2}}{\left|a\right|\Theta}\right)a=\frac{GM}{r^{2}}\end{equation}
}

\noindent \textcolor{black}{Rearranging we get}

\noindent \textcolor{black}{\begin{equation}
a=\frac{GM}{r^{2}}+\frac{2c^{2}}{\Theta}\hat{a}\end{equation}
}

\noindent \textcolor{black}{where $\hat{a}=a/\left|a\right|$, a unit
vector. As shown in {[}38{]}, this formula implies that, even if the
gravitational mass in its vicinity is zero, MiHsC (the second term
on the right hand side) predicts that an object must still accelerate:
there is a minimum acceleration in nature. It is interesting that
this is close to the observed acceleration attributed to dark energy
({[}47{]} and {[}48{]}). Now considering smaller and smaller galaxies,
the mass M in Eq. 4 will decrease and the extra acceleration due to
the new second term will become ever more important. The extra acceleration
is inwards (in the direction of a, ie: â) so the apparent proportion
of dark matter in the galaxy will seem to increase (term 2 divided
by term 1). When M$\rightarrow$0 the MiHsC acceleration will still
be finite, as follows}

\noindent \textcolor{black}{\begin{equation}
a=\frac{2c^{2}}{\Theta}\hat{a}\end{equation}
}

\noindent \textcolor{black}{If we are unaware of this extra MiHsC
term, then the residual observed acceleration will be misinterpreted
as being due to an apparent (or dark) mass $M_{dark}$ so that}

\noindent \textcolor{black}{\begin{equation}
a=\frac{GM_{dark}}{r^{2}}=\frac{2c^{2}}{\Theta}\end{equation}
}

\noindent \textcolor{black}{This apparent dark mass, within a radius
of r=500 parsecs, is predicted by Eq. 6 to be}

\noindent \textcolor{black}{\begin{equation}
M_{dark}=\frac{2c^{2}r^{2}}{G\Theta}=2.3\times10^{39}kg=1.1\times10^{9}M_{\odot}\end{equation}
}

\noindent \textcolor{black}{This agrees with the minimum mass of disc
galaxies of about $10^{9}M_{\odot}$ observed by {[}46{]}. If MiHsC
is right then it implies that these small galaxies have remained above
the MiHsC minimum acceleration by spinning }\textbf{\textcolor{black}{as
if}}\textcolor{black}{ they contained the apparent mass predicted
by Eq. 7. The apparent dark matter in larger galaxies may have a similar
cause, but this is a far more complex problem and outside the scope
of this paper.}
\end{singlespace}

\begin{singlespace}

\section{\noindent \textcolor{black}{Discussion}}
\end{singlespace}

\begin{singlespace}
\noindent \textcolor{black}{To explain in a more intuitive way: as
we consider smaller and smaller galaxies the rotational acceleration
reduces, eventually for very small galaxies the rotational acceleration
is so small that the effect of MiHsC begins to reduce the stars' inertia
and this has the effect of increasing the rotation again. Since the
stars have a very low inertial mass they can easily be bent into rotation
by even a tiny amount of baryonic mass. A balance is reached at the
minimum allowed rotational acceleration. If we are unaware of MiHsC
then we interpret the residual rotation as being due to extra dark
matter. The mass at the balance point has a definite value and has
been predicted above. The fact that this agrees with the observed
mass cutoff for disc galaxies is encouraging.}

\noindent \textcolor{black}{The behaviour of elliptical galaxies has
not been considered here since they are pressure supported and so
Eq. 2 would be more complex, with an extra pressure term. According
to {[}46{]} they can attain masses as low as $10^{7}M_{\odot}$. MiHsC
can explain this qualitatively, since the extra non-rotational accelerations
in these systems make it possible for them to stay above the minimum
acceleration, even with lower rotational accelerations.}

\noindent \textcolor{black}{As a larger-scale independent test, on
a much larger scale, Eq. 7 can be applied to the observable universe
(which has a radius of $\Theta/2$) so that}

\noindent \textcolor{black}{\begin{equation}
M_{dark}=\frac{2c^{2}\left(\frac{\Theta}{2}\right)^{2}}{G\Theta}=\frac{c^{2}\Theta}{2G}=2\times10^{53}kg\end{equation}
}

\noindent \textcolor{black}{This is the apparent total (dark) mass
that the observable universe must have if it is to accelerate fast
enough to satisfy the minimum acceleration of MiHsC. From the Friedmann
cosmological equations ({[}49{]}) the mass that the observable universe
must have to be closed is given by}
\end{singlespace}

\textcolor{black}{\begin{equation}
M_{closed}=\frac{3H^{2}}{8\pi G}\times volume=\frac{3c^{2}}{2\pi G\Theta^{2}}\times\frac{4\pi r^{3}}{3}=\frac{c^{2}\Theta}{4G}\end{equation}
}

\begin{singlespace}
\noindent \textcolor{black}{This mass is $1\times10^{53}kg$ and the
mass of the observable universe including dark matter is thought to
be close to this, although the reason is unknown (this is called the
age, or flatness, problem). As shown above $M_{closed}$ is close
to the minimum mass predicted by MiHsC, so MiHsC could provide an
explanation for the flatness problem.}

\noindent \textcolor{black}{The MiHsC model has a number of problems,
among these are: 1) there is no proven reason why bound and unbound
trajectories should behave differently, 2) how do such long Unruh
waves interact with matter and the Hubble scale? 3) why is it that
the large particle accelerations within stars, and within atoms, do
not need to be considered? 4) The modified inertia predicts dynamical
effects due to both an increased sensitivity to external force, and
changes in momentum and both of these need to be considered.}
\end{singlespace}

\section{\textcolor{black}{A suggested practical test}}

\begin{singlespace}
\noindent \textcolor{black}{Since astronomical tests as discussed
above can be ambiguous, it is important to suggest a direct controllable
experimental test. This is attempted here. The wavelength of the Unruh
radiation seen by accelerated objects {[}35{]} is given by}
\end{singlespace}

\textcolor{black}{\begin{equation}
\lambda=\frac{4\beta\pi^{2}c^{2}}{a}\end{equation}
}

\begin{singlespace}
\noindent \textcolor{black}{For terrestrial accelerations these wavelengths
are too long to be detectable, or to be generated. For example an
object accelerated at $9.8m/s^{2},$ sees Unruh radiation with $\lambda\sim10^{16}m$.
It is suggested here that the first assumption of MiHsC (inertia is
due to Unruh radiation) could be tested by accelerating a particle
around, say, the 1 km ring at the CERN particle accelerator. If the
particle's speed is 0.9c then its acceleration would be $7.3\times10^{13}m/s^{2}$.
Since its acceleration is so high, the Unruh radiation it sees would
be short enough (9.7 km) to be produced artificially (these are long
radio waves). Some extra man-made Unruh radiation could now be applied
to the particle. Since the particle is travelling at 0.9c, and because
of special relativity, a radiation of wavelength 22 km would have
to be used so that the moving particle would see a wavelength of 9.7
km. Then, if assumption 1 of MiHsC is right, the particle would see
more Unruh radiation corresponding to its acceleration, its inertial
mass would increase, and this change in inertia could be detectable
in its trajectory.}
\end{singlespace}

\begin{singlespace}

\section{\noindent \textcolor{black}{Conclusions}}
\end{singlespace}

\begin{singlespace}
\noindent \textcolor{black}{The cosmic acceleration attributed to
dark energy, the observed minimum disc galaxy mass of about $10^{9}M_{\odot}$
and the Hubble mass, can be predicted using a model that assumes that
1) inertia is due to Unruh radiation, and 2) this radiation is subject
to a Hubble-scale Casimir effect.}

\noindent \textcolor{black}{It is proposed that assumption 1 of MiHsC
can be tested using a particle accelerator to accelerate particles
to the extreme, so that the Unruh radiation they see (and that may
determine their inertial mass) is short enough to be supplemented
using man-made radiation. This may allow control of the particles'
inertia, with detectable consequences.}
\end{singlespace}

\begin{singlespace}

\section*{\noindent \textcolor{black}{Acknowledgements}}
\end{singlespace}

\begin{singlespace}
\noindent \textcolor{black}{Many thanks to M.Bate, J.R.Sambles, D.Price,
R.Woodard, K.A.Rosser, B.Kim and an anonymous reviewer for advice
and support.}
\end{singlespace}

\begin{singlespace}

\section*{\noindent \textcolor{black}{References}}
\end{singlespace}

\begin{singlespace}
\noindent \textcolor{black}{{[}1{]} Zwicky, F., 1933. Der Rotverschiebung
von extragalaktischen Nebeln. }\textit{\textcolor{black}{Helv. Phys.
Acta}}\textcolor{black}{, 6, 110.}

\noindent \textcolor{black}{{[}2{]} Rubin, V., 1983. Science. 220,
1339.}
\end{singlespace}

\textcolor{black}{{[}3{]} Bosma, A., 1981. AJ, 86, 1791.}

\textcolor{black}{{[}4{]} Rubin, V., W. Ford, N. Thonnard, D. Burstein,
1982. ApJ, 261, 439.}

\begin{singlespace}
\noindent \textcolor{black}{{[}5{]} Bernabei, R. Presentation at BEYOND
conference, 1-6 February, 2010, Cape Town, S. Africa. arXiv:1002.1028.}

\noindent \textcolor{black}{{[}6{]} Ahmed, Z., et al., 2009. CDMS
collaboration, Phys. Rev. Lett., 102, 011301.}

\noindent \textcolor{black}{{[}7{]} XENON10 collaboration, 2009. Phys.
Rev. D., 80, 115005.}

\noindent \textcolor{black}{{[}8{]} Khlopov, M.Yu., AIP proceedings
of the invisible universe international conference, UNESCO-Paris,
June 29-July 3, 2009. arXiv:0911.5685.}

\noindent \textcolor{black}{{[}9{]} Milgrom M., 1983. A modification
of the Newtonian dynamics as a possible alternative to the hidden
mass hypothesis. $Astrophysical~Journal$, 270, 365.}

\noindent \textcolor{black}{{[}10{]} Iorio, L., 2008. Constraining
MOND with Solar System Dynamics, Journal of Gravitational Physics,
vol. 2, no.1, pp. 26-32, 2008.}

\noindent \textcolor{black}{{[}11{]} McGaugh, S.S., 2007. Science,
Letters, 318, 568.}

\noindent \textcolor{black}{{[}12{]} Anderson, J.D., P.A. Laing, E.L.
Lau, A.S. Liu, M.M. Nieto, S.G. Turyshev, 1998. Phys. Rev. Lett. 81,
2858.}

\noindent \textcolor{black}{{[}13{]} Anderson, J.D, P.A. Laing, E.L.
Lau, A.S. Liu, M.M. Nieto, S.G. Turyshev, 2002. Phys. Rev. D., 65,
082004.}

\noindent \textcolor{black}{{[}14{]} Pitjeva, E.V., 2006. Limitations
on some physical parameters from position observations of planets.
Paper presented at the 26th meeting of the IAU, 22-23 August 2006,
Prague, Czech Republic, Joint discussion 16, No. 55.}

\noindent \textcolor{black}{{[}15{]} Iorio, L, G. Giudice, 2006. What
do the orbital motions of the outer planets of the Solar System tell
us about the Pioneer anomaly. New Astronomy, 11, 8, 600-607.}

\noindent \textcolor{black}{{[}16{]} Iorio, L., 2007a. Can the Pioneer
anomaly be of gravitational origin? A phenomenological answer. Foundations
of Physics, 37, 6, 897-918.}

\noindent \textcolor{black}{{[}17{]} Iorio, L., 2007b. Jupiter, Saturn
and the Pioneer anomaly: a planetary based independent test. J. Grav.
Phys., 1, 1, 5-8.}

\noindent \textcolor{black}{{[}18{]} Iorio, L., 2010. Does the Neptunian
system of satellites challenge a gravitational origin for the Pioneer
anomaly? MNRAS, doi:10.1111/j.1365-2966.2010.16637.x}

\noindent \textcolor{black}{{[}19{]} Standish, E.M., 2008. Planetary
and lunar ephemerides: testing alternate gravitational theories. In
Macias, A., Lammerzahl, C. Camcho, A., eds, AIP Conference Proceedings:
Recent developments in gravitation and cosmology - 3rd Mexican Meeting
on Mathematical and experimental physics. 977, AIP, p. 254.}

\noindent \textcolor{black}{{[}20{]} Standish, E.M., 2010. Testing
alternate gravitational theories. Proceedings of the IAU symposium
261, 179-182.}

\noindent \textcolor{black}{{[}21{]} Iorio, 2009. Can the Pioner anomaly
be induced by velocity-dependent forces? Tests in the outer regions
of the Solar System with planetary dynamics. Int. J. Modern Physics,
18, 6, 94-958.}

\noindent \textcolor{black}{{[}22{]} Page, G.L., J.F. Wallin, D.S.
Dixon, 2009. How well do we know the orbits of the outer planets?
ApJ, 697, 1226.}

\noindent \textcolor{black}{{[}23{]} Page, G.L., D.S. Dixon, J.F.
Wallin, 2006. Can minor planets be used to assess gravity in the outer
Solar system. ApJ, 642, 606.}

\noindent \textcolor{black}{{[}24{]} Wallin, J.F., D.S. Dixon, G.L.
Page, 2007. Testing gravity in the outer solar system: results from
trans-Neptunian objects. ApJ, 666, 1296.}

\noindent \textcolor{black}{{[}25{]} Bertolami, O., Francisco, F.,
Gil, P.J.S., Paramos, J., 2008. Thermal analysis of the Pioneer anomaly:
a method to estimate radiative momntum transfer.}

\noindent \textcolor{black}{{[}26{]} Rievers, B., C. Lammerzahl, H-J.
Dittus, 2009. New J. Phys., 11, 113032, 24pp.}

\noindent \textcolor{black}{{[}27{]} Rievers, B., S. Bremer, M. List,
C. Lammerzahl, H-J, Dittus, 2010. Thermal dissipation force modelling
with preliminary results for Pioneer 10/11. Acta Astronautica, 66,
3-4, 467-476.}

\noindent \textcolor{black}{{[}28{]} Toth, V.T., S.G. Turyshev, 2009.
Thermal recoil force, telemetry and the Pioneer anomaly. Phys. Rev.
D., 79, 043011.}

\noindent \textcolor{black}{{[}29{]} Brownstein, J.R., J.W. Moffat,
2006. Class. Quantum Gravity, 23, 3427.}
\end{singlespace}

\textcolor{black}{{[}30{]} Gerrard, M.B. and T.J. Sumner, 2008. arXiv:0807.3158.}

\begin{singlespace}
\noindent \textcolor{black}{{[}31{]} Turyshev, S.G., V.T. Toth, 2010.
arXiv:1001.3686.}

\noindent \textcolor{black}{{[}32{]} Rathke, A., D. Izzo, 2006. Options
for a non-dedicated mission to test the Pioneer anomaly. J. Spacecraft
Rockets, 43, 806.}

\noindent \textcolor{black}{{[}33{]} Milgrom, M., 1999. Phys. Lett.
A, 253, 273.}

\noindent \textcolor{black}{{[}34{]} Hawking, S., 1974. Nature, 248,
30.}

\noindent \textcolor{black}{{[}35{]} Unruh, W.G., 1976. Phys. Rev
D., 14, 870.}

\noindent \textcolor{black}{{[}36{]} Haisch, B., A. Rueda, H.E. Puthoff,
1994. Phys. Rev. A., 49, 678.}

\noindent \textcolor{black}{{[}37{]} Milgrom, M., 1994. Ann. Phys.,
229, 384.}

\noindent \textcolor{black}{{[}38{]} McCulloch, M.E., 2007. Modelling
the Pioneer anomaly as modified inertia. $MNRAS$, 376, 338-342. arXiv:astro-ph/0612599.}

\noindent \textcolor{black}{{[}39{]} Freedman, W.L., 2001. ApJ, 553,
47.}

\noindent \textcolor{black}{{[}40{]} McCulloch, M.E., 2008. Modelling
the flyby anomalies using a modification of inertia. }\textit{\textcolor{black}{MNRAS-letters}}\textcolor{black}{,
389(1), L57-60. arXiv:0806.4159}

\noindent \textcolor{black}{{[}41{]} Anderson, J.D, J.K. Campbell,
J.E. Ekelund, J. Ellis and J.F. Jordan, 2008. Anomalous orbital energy
changes observed during spacecraft flybys of Earth. }\textit{\textcolor{black}{Phys.
Rev. Lett}}\textcolor{black}{., 100 (091102)}

\noindent \textcolor{black}{{[}42{]} Nieto, M.M., J.D.Anderson, 2009.
Earth flyby anomalies, Physics Today, 62, No. 10, 76-77.}

\noindent \textcolor{black}{{[}43{]} Iorio L., 2009. The effect of
general relativity on hyperbolic orbits and its application to the
flyby anomaly, Scholarly Research Exchange, Volume 2009. Article ID
807695.}

\noindent \textcolor{black}{{[}44{]} De Lorenzi, V.A., M. Faundez-Abans
and J.P. Pereira, 2009. A \& A, 503, L1.}

\noindent \textcolor{black}{{[}45{]} McCulloch, M.E., 2010. Can the
Tajmar effect be explained as a modification of inertia? EPL, 89,
19001. arXiv:0912.1108}

\noindent \textcolor{black}{{[}46{]} McGaugh, S.S., J.M. Schombert,
W.J.G de Blok and M.J. Zagursky, 2009. The baryon content of cosmic
structures. Astrophys. J., 708, L14-17.}

\noindent \textcolor{black}{{[}47{]} Perlmutter, S., et al., 1999.
The supernovae cosmology project {}``Measurement of Omega and Lambda
from 42 high redshift supernovae.''. Astrophysical Journal, 517,
565-586.}

\noindent \textcolor{black}{{[}48{]} Riess, A.G., 1998. Observational
evidence from supernovae for an accelerating universe and a cosmological
constant. Astronomical Journal., 116, 1009-1038.}

\noindent \textcolor{black}{{[}49{]} Coles, P. and F. Lucchin, 2002.
Cosmology: the origin and evolution of cosmic structure (2nd ed.).
J.Wiley and Sons Ltd.}\end{singlespace}

\end{document}